\documentclass[prb,twocolumn, notitlepage,showpacs,preprintnumbers,amsmath,amssymb,10pt,longbibliography]{revtex4-1}

\usepackage{graphicx}
\usepackage{amsmath}
\usepackage{physics}
\usepackage{hyperref}
\usepackage{amsbsy}
\usepackage{bm}

\newcommand{\omegat}{\tilde{\omega}}
\newcommand{\ham}{\widetilde{\mathcal{H}}}

\newcommand{\sigmabar}{\bar{\sigma}}
\newcommand{\fl}[1]{(#1)} 

\renewcommand{\Re}{\operatorname{Re}}
\renewcommand{\Im}{\operatorname{Im}}

\newcommand{\mat}[1]{#1}

\begin{document}

\title{Theoretical study of impurity-induced magnetism in FeSe}
\author{Johannes H. J. Martiny,$^1$ Andreas Kreisel,$^2$ and Brian M. Andersen$^3$}
\affiliation{$^1$Center for Nanostructured Graphene (CNG), Dept. of Micro- and Nanotechnology, Technical University of Denmark, DK-2800 Kongens Lyngby, Denmark}
\affiliation{$^2$Institut f\" ur Theoretische Physik, Universit\" at Leipzig, D-04103 Leipzig, Germany}
\affiliation{$^3$Niels Bohr Institute, University of Copenhagen, Lyngbyvej 2, DK-2100 Copenhagen, Denmark}

\date{November 5, 2018}

\begin{abstract}
Experimental evidence suggests that FeSe is close to a magnetic instability, and recent scanning tunneling microscopy (STM) measurements on FeSe multilayer films have revealed stripe order locally pinned near defect sites. Motivated by these findings, we perform a theoretical study of locally induced magnetic order near nonmagnetic impurities in a model relevant for FeSe. We find that relatively weak repulsive impurities indeed are capable of generating short-range magnetism, and explain the driving mechanism for the local order by resonant $e_g$-orbital states. In addition, we investigate the importance of orbital-selective self-energy effects relevant for Hund's metals, and show how the structure of the induced magnetization cloud gets modified by orbital selectivity. Finally, we make concrete connection to STM measurements of iron-based superconductors by symmetry arguments of the induced magnetic order, and the basic properties of the Fe Wannier functions relevant for tunneling spectroscopy.

\end{abstract}
\maketitle
\section{Introduction}

The understanding of the electronic properties of the material FeSe continues to pose an interesting challenge to the research community of iron-based superconductors. Controversial current topics include the reasons for its  modified electronic structure (compared to DFT calculations), the nature of the nematic phase, and the origin of the highly anisotropic superconducting gap structure.\cite{BoehmerKreisel_review} There is considerable interest in resolving these issues both for our general understanding of correlated superconductors in general, and FeSe in particular due to the ability to significantly enhance its superconducting transition temperature $T_c$ by pressure, intercalation, or dosing.\cite{Medvedev2009,Bendele2012,Sun_2016,Kothapalli2016,Burrard-Lucas2013,Ying2011II,Krzton-Maziopa2016} In addition, while bulk FeSe exhibits a $T_c\sim 9$ K, a single monolayer of FeSe on STO has been shown to superconduct up to $\sim 65$ K.\cite{Wang12} On the other hand, thicker films suppress  superconductivity and exhibit a strong nematic phase for reasons that remain unclear at present.\cite{ZhangY15}

A striking difference between FeSe and most of the iron-based superconductors is the lack of magnetic ordering in FeSe. Even though the tetragonal to orthorhombic transition takes place around $T_s\sim 90$ K, there is no evidence for long-range static magnetic order setting in at lower temperatures. However, there is experimental evidence that FeSe is close to a magnetic instability at low temperatures, as seen by the diverging spin-lattice relaxation rate $1/T_1T$ versus $T$ by NMR.\cite{Imai2009} For example, modest pressures exceeding $\sim 0.8$ GPa induce static stripe antiferromagnetism indicating that FeSe at ambient pressure is parametrically close to the ordered magnetic phase.\cite{Medvedev2009,Bendele2012,Sun_2016,Kothapalli2016} The resulting temperature-pressure phase diagram describing the pressure dependence of nematic, magnetic, and superconducting orders has been recently described theoretically in terms of pressure-dependent electronic interactions.\cite{Scherer_FeSe} The importance of low-energy magnetic fluctuations in FeSe (at ambient pressure) has been also pointed out by recent inelastic neutron scattering experiments revealing a rich temperature and momentum dependence of the scattering intensity.\cite{Rahn2015,Wang2015GS,Wang2015,Ma_2017} As a function of temperature, spectral weight is shifted from N\'{e}el-like fluctuations to stripe-like $(\pi,0)$ fluctuations. Thus at low temperatures the magnetic fluctuations at low energies are entirely dominated by the stripe-like fluctuations.

The proximity to a stripe magnetic instability suggests the possibility of disorder-induced magnetism in FeSe. Naively various imperfections such as impurities and twin boundaries may relatively easily induce weak local magnetic order by the presence of a nearby magnetic quantum critical point.\cite{Andersen2010} Despite the fact that very high quality FeSe crystals can be made\cite{Boehmer2013}, and disorder-generated magnetism does not appear to be widespread in those samples, a number of recent experimental results do find evidence of local magnetism. For example, the close similarity in the behavior of the magnetostriction and uniform susceptibility between BaFe$_2$As$_2$ and FeSe in the nematic phase, led He \textit{et al.}\cite{He2018} to propose that short-range stripe magnetic order exists in FeSe. Evidence of dilute static magnetism possibly arising from impurities has also been recently put forward by $\mu$SR measurements on high quality single crystals.\cite{Grinenko2018} Earlier $\mu$SR studies of FeSe$_{0.85}$ also found evidence of a dilute and randomly distributed static magnetic signal.\cite{Khasanov2008} Related to these findings, an STM study of FeSe multilayer films found clear evidence of charge stripe order centered near Fe vacancy sites.\cite{Li2017} This study reveals a clear example of impurity-induced local order, and it was suggested by the authors that the observed charge stripes are the natural associated charge modulations induced by the magnetic fluctuations pinned by the defect sites.\cite{Li2017} The presence of disorder-pinned antiferromagnetic order was also recently suggested to be at play in parent as-grown films of FeSe on STO.\cite{Zhou2018} Finally, we point out recent NMR studies of FeSe finding evidence of static short-range nematic order above $T_s$.\cite{Wang2017III_pub,Wiecki2017} It remains an interesting question if and how local magnetic order may be connected to these NMR observations.

From a theoretical perspective, the low-energy magnetic fluctuations in bulk FeSe have been described within an itinerant approach which successfully captured the temperature and momentum dependence of the spin excitations.\cite{Mukherjee2015,Kreisel15,Kreisel_2018arXiv180709482K} However, this is only true if one includes so-called orbital-selective effects in the theory, i.e. the fact that distinct orbitals experience different self-energy renormalizations leading to orbital-dependent mass enhancements and quasi-particle weights.\cite{Haule2009,Yin2011,deMedici2011,Lanata2013,Georges2013,deMedici2014,Fanfarillo2015,deMedici2017,Biermann_review,Guterding2017,deMedici_review} These properties are characteristics of Hund's metals, and agree with recent STM quasi-particle interference measurements both in the normal state and superconducting phases.\cite{Kostin2018,Sprau2017,Kreisel2017} 

In terms of impurity-physics in unconventional superconducting materials, a number of theoretical works have pointed out the interesting role of electronic interactions in dressing bare impurity potentials.\cite{Alloul_RMP2009,Tsuchiura2001,Wang2002,Zhu2002,Chen2004,Harter2007,Kontani2006, Navarro_LiFeAs,Gastiasoro2015_NovelMag,Gastiasoro2016}. In addition, there are nontrivial effects from the multi-band electronic structure of this family of materials. For example, in the nematic state, nonmagnetic disorder may lead to short-range anisotropic magnetic order which has been proposed to explain unusual transport phenomena is Co-doped BaFe$_2$As$_2$.\cite{Gastiasoro2014} Regarding the superconducting state, there are also novel suggested impurity effects including disorder-enhanced $T_c$ due to local density of states (LDOS) enhancements from bound states generated in off-Fermi level bands.\cite{Gastiasoro2017_2}

In this paper, we combine realistic microscopic modeling of FeSe with impurity studies to address the role of local nucleated short-range magnetic order in this material. We apply the so-called Chebyshev Bogoliubov-de Gennes method to study large real-space systems, and map out the phase diagram of local magnetic order as a function of onsite Coulomb repulsion $U$ and impurity potential $V_0$. We find a favorable impurity potential range for induced local order. In addition, we discuss the role of orbital selectively in the self-consistency equations, and show how the associated self-energy effects are directly tied to the local internal structure of the induced magnetization clouds surrounding impurity sites with favorable potentials able to generate induced order. We suggest that the experimental evidence of local magnetism in FeSe may be caused by a particular class of disorder in this material.

\section{Method}
We start from a fitted tight binding model for the nematic phase of FeSe
with the Hamiltonian
\begin{align}
\mathcal{H}_{0} = \sum_{ij,  \mu \nu , \sigma} (t^{\mu \nu}_{ij}- \delta_{ij}\delta_{\mu \nu} \mu_0) c^\dagger_{i\mu \sigma} c_{j\nu \sigma} + H.c., \label{eq:H_0}
\end{align}
where $\mu, \nu$ span the d-orbitals of the two inequivalent iron atoms in the unit cell, and $t^{\mu \nu}_{ij}$ denote the hopping elements detailed in Ref.~\onlinecite{Kreisel2017}. A point-like impurity at a site $i'$ is described by the term 
\begin{align}
\mathcal{H}_{imp} &= V_{0} \sum_{\mu, \sigma} c^{\dagger}_{i' \mu \sigma } c_{i' \mu \sigma}, 
\end{align} 
where the sum now spans the orbitals of a single iron site. Interactions are initially included using the usual multiorbital Hubbard-Hund model 
\begin{align}
 \mathcal{H}_{int} &=~U \sum_{i,\mu} n_{i\mu \uparrow} n_{i\mu \downarrow}  
 +U' \sideset{}{'}\sum_{i,\mu < \nu,\sigma} n_{i\mu \sigma}n_{i\nu \bar{\sigma}}  \\
 &~~~+(U'-J)\sideset{}{'}\sum_{i, \mu < \nu, \sigma}n_{i\mu \sigma}n_{i\nu \sigma} \nonumber \\
 &~~~+J \sideset{}{'}\sum_{i, \mu < \nu, \sigma} c_{i\mu \sigma}^\dagger c_{i\nu \bar{\sigma}}^\dagger c_{i\mu \bar{\sigma}} c_{i\nu \sigma} \nonumber \\
 &~~~+J' \sideset{}{'}\sum_{i, \mu \neq \nu} c_{i\mu \uparrow}^\dagger c_{i\mu \downarrow }^\dagger c_{i\nu \downarrow }c_{i\nu \uparrow}, \nonumber
 \label{eq:H_int}
\end{align}
where we set $J= J'= U/4$ and use spin-rotational invariant interactions, $U'=U-2J$, and the sums $\sum^\prime$ only give a contribution
when the indices $\mu$ and $\nu$ label an orbital on the same iron atom.

The interactions are included at the mean field level, yielding the mean field Hamiltonian 
\begin{widetext}
\begin{align}
 \mathcal{H}_{int}^{MF} &= \sum_{i,\nu, \sigma} \Big\lbrack U\ev*{n_{i\nu \sigmabar}} +\sideset{}{'}\sum_{\mu \neq \nu}\bigl\{U' \ev*{n_{i\mu \sigmabar}}+(U'-J)\ev*{n_{i\mu \sigma}} \bigr\}\Big\rbrack c^\dagger_{i\nu \sigma}c_{i\nu \sigma} \\
 &~~~ - \sideset{}{'}\sum_{i,\mu \neq \nu, \sigma} \Big\lbrack (U'-J) \ev*{c^\dagger_{i\nu \sigma}c_{i\mu \sigma}}  - J' \ev*{c^\dagger_{i\mu \sigmabar}c_{i\nu \sigmabar}} - J \ev*{c^\dagger_{i\nu \sigmabar}c_{i\mu \sigmabar}} \Big\rbrack c^\dagger_{i\mu \sigma} c_{i\nu \sigma}  \nonumber \\
 &~~~ -\sum_{i,\nu, \sigma } \Big\lbrack U\ev*{c^\dagger_{i\nu \sigma}c_{i\nu \sigmabar}}
+J \sideset{}{'}\sum_{\mu \neq \nu} \ev*{c^\dagger_{i\mu \sigma}c_{i\mu \sigmabar}}\Big\rbrack c^\dagger_{i\nu \sigmabar} c_{i\nu \sigma}  
  - \sideset{}{'}\sum_{i,\mu \neq \nu, \sigma} \Big\lbrack U'\ev*{c^\dagger_{i\nu \sigma}c_{i\mu \sigmabar}}
+J'\ev*{c^\dagger_{i\mu \sigma}c_{i\nu \sigmabar}}\Big\rbrack c^\dagger_{i\mu \sigmabar} c_{i\nu \sigma}.  \nonumber \label{eq_MF}
\end{align} 
\end{widetext}

The Hamiltonian $\mathcal{H}=\mathcal{H}_{0}+\mathcal{H}_{imp}+\mathcal{H}_{int}^{MF}$ defines the ``bare'' version of our model, where effects of orbital selectivity (discussed further below) are not included. The results derived from this bare model will serve as a comparison basis for another model defined below which includes the effects of orbital selectivity. 

Unrestricted self-consistent calculations of the density and magnetization mean fields for the tight binding models are performed using the Chebyshev Bogoliubov-de Gennes (CBdG) method,\cite{Weisse_KPM} wherein the electronic Greens function of a Hamiltonian $\mathcal{H}$ is expanded in a series of orthogonal polynomials. We will provide a brief outline of this procedure below. The starting point of the expansion procedure is the estimation of extremal eigenvalues  $E_{\text{min}}, E_{\text{max}}$ which are obtained by explicitly diagonalizing the Hamiltonian in a small system. We can then define the rescaled Hamiltonian 
\begin{align}
\ham = (\mathcal{H}-b)/a,
\end{align}
with $b=(E_{\text{max}}+E_{\text{min}})/2$ and $a=(E_{\text{max}}-E_{\text{min}})/(2-\delta)$, where $\delta = 0.001$ is a small parameter introduced to avoid divergence at the edges of the domain. The rescaled Hamiltonian $\ham$ has eigenvalues in the interval $ (-1,1)$, which coincides with the domain of the Chebyshev polynomials.   

Defining the rescaled energy $\omegat = (\omega-b)/a \in (-1,1)$, the Greens function can then be expanded as 
\begin{align}
G_{\mu \nu}^{\sigma \sigma'}(i,j,\omegat) &= \lim_{\eta \to 0}\mel{c_{i \mu \sigma}}{\frac{1}{\omegat + i \eta - \ham}}{c_{j \nu \sigma'}^\dagger}  \\
&= \frac{-2i}{\sqrt{1-\omegat^2}}\sum_{n=0}^{N-1} a_{\mu \nu,n}^{\sigma \sigma'}(i,j)\exp(-in \arccos(\omegat)) \nonumber, \label{eq:GF_exp}
\end{align}
with $\ket*{c_{j\nu \sigma}^\dagger} = c_{j \nu \sigma}^\dagger \ket{0}$, and expansion coefficients  
\begin{align}
a_{\mu \nu, n}^{\sigma \sigma'}(i,j) &= \frac{1}{1+\delta_{0,n}} \mel{c_{i \mu \sigma}}{T_n(\ham)}{c_{j \nu \sigma'}^{\dagger}}
\end{align}
where $T_n$ is the $n$th Chebyshev polynomial of the first kind. The problem has therefore been reduced to finding the expansion coefficients, which are obtained using the recursion relation of the Chebyshev polynomials. Defining the intermediate states $\ket{j_n} = T_n(\ham) \ket*{c_{j \nu \sigma'}^\dagger}$, we can generate coefficients recursively starting from an initial state
\begin{subequations}
\begin{align}
\ket{j_0} &= \ket*{c_{j \nu \sigma'}^\dagger}, \\
\ket{j_1} &= \ham \ket*{c_{j \nu \sigma'}^\dagger}, \\
\ket{j_{n+1}} &= 2\ham \ket{j_n}-\ket{j_{n-1}}.
\end{align}
\end{subequations}
The full expansion coefficients can be then obtained as the inner product $a_{\mu \nu,n}^{\sigma \sigma'}(i,j) = \braket*{c_{i \mu \sigma} }{ j_n}$. An artificial broadening of $\eta = 1$ meV is included in the Greens function by applying the Lorentz kernel during the expansion.\cite{Weisse_KPM}  

The mean fields in Eq. (\ref{eq_MF}) and the local density of states (LDOS) then follow from 
\begin{align}
\ev{c^{\dagger}_{i\mu \sigma}c_{i\nu \sigma'}} &= \int_{-1}^1 \dd{\tilde{\omega}} \Im G_{\mu \nu}^{\sigma \sigma'}(i,i,\tilde{\omega})f(\tilde{\omega}),\label{eq:mean_fields}\\
\rho^{\sigma}_{\mu}(i,\omega) &= -\frac{1}{\pi}  \Im G_{\mu \mu}^{\sigma \sigma}(i, i, \omega), \label{eq:LDOS}
\end{align}
with $f(\omegat)$ the Fermi-Dirac distribution function which is evaluated at a temperature of $1K$ in all following calculations. (For the study of FeSe below this implies that we are deep within the nematic phase in the undoped system.) The energy integrals for the mean fields can be obtained efficiently using Chebyshev-Gauss quadrature in a similar fashion as Ref. \onlinecite{Weisse_KPM}, leaving the sparse matrix-vector products as the limiting part of the full calculation. We find that these mean fields are converged at $N = 1000$ expansion coefficients, and use this value for all calculations apart from when we plot the LDOS at high energy resolution (then, $N = 20000$). In agreement with Ref. \onlinecite{Weisse_KPM}, we find that this procedure is extremely efficient for selfconsistent calculations in large multi-orbital systems such as our considered ten orbital model, while yielding results consistent with the conventional BdG method. We stress that all calculations below are fully unrestricted in all orbitals and sites.

\section{Results}
\begin{figure}[tb!]
\includegraphics[width = \linewidth]{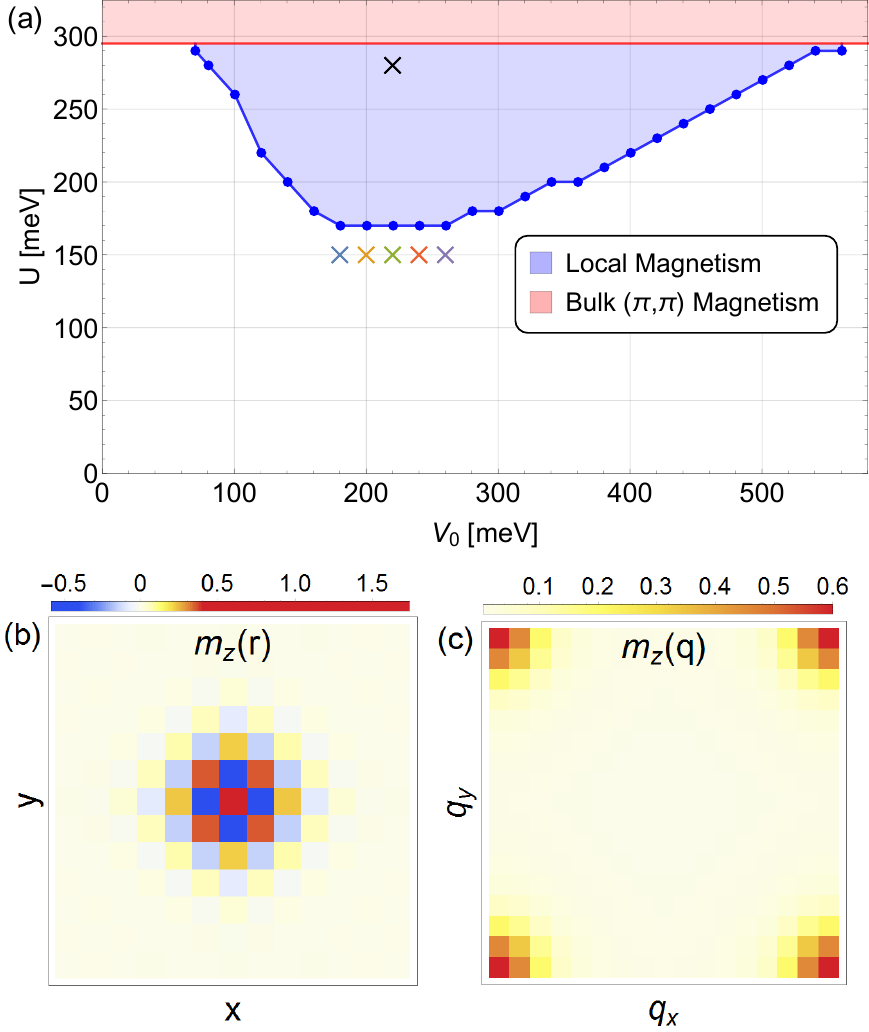}
	\caption{
	\fl{a} Phase diagram of impurity-induced magnetization as a function of the impurity potential strength $V_0$ and Hubbard $U$. The phase diagram shows the bulk $(\pi, \pi)$ phase above $U_c =  295$meV (red), and a region of impurity nucleated local magnetic order (blue) just below the bulk order. The system is most susceptible to the formation of local magnetic order for impurity potentials close to $V_0 \approx 220$meV. 
	\fl{b} Magnetization for $V_0 = 220$meV, $U = 280$meV deep in the pocket of local magnetic order (indicated by the black cross in the phase diagram \fl{a}), alongside \fl{c} the Fourier transform showing the local $(\pi, \pi)$ order. 
 	 }
 	 \label{fig:PD_0}
	
\end{figure}

The phase diagram of magnetization versus $V_0$ and $U$ is obtained by initializing a $12 \times 12$ system with a central impurity surrounded by a small uniformly spin polarized region, and then converging the mean fields for given Hubbard $U$ and impurity potentials $V_0$. Convergence is defined as a maximal variation of the set of mean fields $n$ of Eq. (\ref{eq:mean_fields}) of $\max(n_{m-1} - n_m) < 10^{-7}$ between iteration steps $m-1$ and $m$, for at least $100$ iteration steps. This is usually accomplished within 1500 iteration steps of the CBdG procedure. Consistent with previous susceptibility calculations,\cite{Kreisel2017} we find that the homogeneous system ($V_0 = 0$) displays 
a transition to a global $(\pi,\pi)$ antiferromagnetic phase at a critical $U_c = 295$meV. 
Approaching this transition from below, we find the possibility of local magnetic order nucleated by the impurity site for a range of potentials $V_0$ displayed in Fig. \ref{fig:PD_0} \fl{a}. At the phase boundary the local order sets in at $V_0 = 70$meV, and extends until $V_0 = 560$meV, with only repulsive potentials being able to induce local magnetic order. 
The local magnetic structure inside this phase mirrors the bulk $(\pi, \pi)$ phase as demonstrated in Fig. \ref{fig:PD_0}\fl{b-c}, displaying the real space magnetization (b) and its Fourier transform (c). The orbital splitting included in $\mathcal{H}_{0}$ to describe the nematic order of FeSe at low $T$, induces a negligible degree of $C_4$-symmetry breaking in the magnetization, and hence the results in Fig. \ref{fig:PD_0}\fl{b-c} appear $C_4$ symmetric even though they are, strictly speaking, only $C_2$ symmetric. 

\begin{figure*}[tb]
\includegraphics[width = \linewidth]{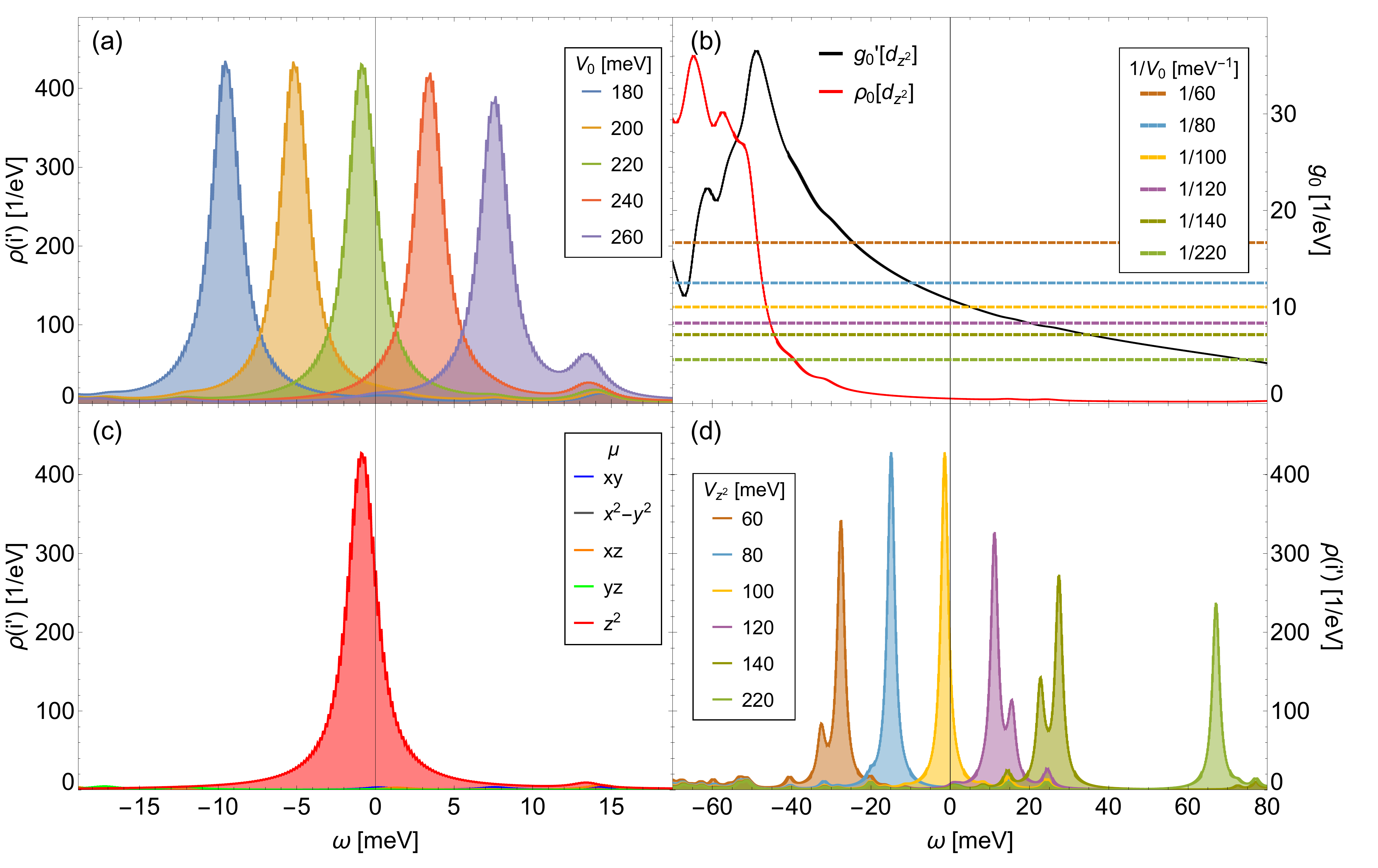}
\caption{
\fl{a} LDOS at the impurity site $i'$ for $U= 150$meV just below the local magnetic transition, displaying a clear progression of bound states with varying $V_0$. As the bound state approaches the Fermi level, the local magnetic transition sets in as a local Stoner transition. 
\fl{b} Real (black line) and imaginary (red line, $\propto LDOS$) part of the $d_{z^2}$ component of the Greens function. Resonant states are expected at energies where the inverse impurity potential (dashed lines) matches the real part of the Greens function while the LDOS is gapped.  
\fl{c} Orbitally resolved LDOS of a single impurity potential $V = 220$meV, demonstrating that the resonant state is almost purely of $d_{z^2}$ character due to the negligible density of states for this orbital. 
 \fl{d} Removing selfconsistency and using a purely $d_{z^2}$ impurity in the tight binding calculation yields resonant states matching the T-matrix solution within the broadening.
   }
\label{fig:BS_variation}
\end{figure*}

We now turn to the underlying reason for stabilization of local magnetic order. Previous studies of impurity-induced magnetization have found a link between local magnetic order and impurity resonant states formed at the Fermi level just below the local magnetic transition.\cite{Gastiasoro2015_NovelMag,GastiasoroPhDthesis} This suggest a mechanism of locally enhanced LDOS providing a local Stoner transition to a magnetic state.\cite{Gastiasoro2015} In Fig. \ref{fig:BS_variation} \fl{a} we investigate this local Stoner scenario. We fix $U = 150$meV, i.e. just below the local magnetic transition and show the LDOS near the Fermi level for varying values of the impurity potential $V_0$, as marked by the line of colored crosses in Fig. \ref{fig:PD_0} \fl{a}. We find that the point in ($U, V_0$)-phase space where the system is most susceptible to local magnetic order, i.e. where the critical coupling line $U_c(V_0)$ has its lowest value, corresponds exactly to the impurity potential where the resonant state crosses the Fermi level ($V_0 \approx 220$meV). This indicates that the onset of local magnetic order can be understood as a local Stoner transition. The role of these resonant states in inducing local magnetism, and enhancing superconductivity, has been recently discussed in Refs. \onlinecite{Gastiasoro2016} and \onlinecite{Gastiasoro17_enhance}.

The emergence of these resonant states can in turn be understood from the real space Greens function in the presence of a point-like impurity at the origin ($r = 0$) 
\begin{align}
\mat{G}(r,\omega) &= \mat{G_0}(r,\omega) + \mat{G_0}(r-0, \omega) \mat{T}(\omega) \mat{G_0}(0-r,\omega),
\end{align}
where each quantity is a matrix containing the spin and orbital components of Eq. (\ref{eq:GF_exp}), and we have defined the impurity T-matrix
\begin{align}
\mat{T}(\omega) = \frac{\mat{V}_0}{\mathbb{I} - \mat{V}_0 \sum_{k} \mat{G}_{0}(k,\omega)} 
=  \frac{\mathbb{I}}{(\mat{V}_0)^{-1} - \mat{g}_0(\omega)},
\end{align}
with the shortened notation $\mat{g_0}(\omega) = \mat{G_0}(0,\omega)$ for the local Greens function in the absence of impurities.
The impurity-induced change in the LDOS can then be defined using Eq. (\ref{eq:LDOS})
\begin{align}
\delta \rho(r,\omega) &= \rho(r,\omega) - \rho_0(r,\omega)\\
&= - \frac{1}{\pi} \Im \left[G_0(r-0,\omega)T(\omega)G_0(0-r,\omega) \right],\notag
\end{align}
from which we see that impurity bound states correspond to poles of the T-matrix. If the impurity and local Greens function are diagonal matrices, we find five independent criteria for the formation of bound states
\begin{align}
\det[(\mat{V}_0)^{-1} - \mat{g}_{0}(\omega)] =  \prod_\mu (\frac{1}{V_{0}} - g_{0}^{\mu}(\omega)) = 0,
\end{align}
i.e. a bound state appears at an energy $\omega = \xi$ if for any orbital $\mu$  
\begin{subequations}
\begin{align}
0 &= -\pi \rho_0^{\mu}(0, \xi),\\
\frac{1}{V_0} &= \Re g_0^{\mu\mu}(\xi). 
\end{align}
\end{subequations}

 Solutions to these equations for any energy $\xi$ correspond to true bound states with impurity site LDOS $\delta \rho(0, \xi) \propto \delta(\xi)$,while resonant states are allowed as complex solutions $\xi= \xi' + i\xi''$ with a broadened Lorentzian shape in the impurity site LDOS.\cite{GastiasoroPhDthesis} 
If we consider a quasi-gapped region where $\rho_0^{\mu}(0, \omega) \approx 0$ for some orbital $\mu$, the T-matrix solution predicts resonant states with orbital character $\mu$, and the resonant state energy $\xi$ determined by the impurity strength $V_0$. 

Fig. \ref{fig:BS_variation} \fl{b} shows the graphical solution to these equations obtained from a converged homogeneous system ($V_{0} = 0, U = 150$meV). The three $t_{2g}$ orbitals all have finite spectral weight at the Fermi level, leaving only the two quasi-gapped $e_g$ orbitals , $d_{z^2}, d_{x^2-y^2}$, as candidates for the resonant state. Of these only the $d_{z^2}$ real part of the Greens function fulfills the second condition in this energy interval. This results in resonant states of purely $d_{z^2}$ character as shown in \fl{c} where each orbital component of the LDOS is plotted. Since the $d_{z^2}$ LDOS is quasi-gapped in an extended energy interval, the location of the resonant state varies smoothly with the impurity potential as evident in \fl{a}. 

We note that while the progression of resonant state energies matches the quasi-gapped region and the slope of the real part of the orbital Greens function, a discrepancy of $\approx 80 $meV in the exact position of the resonant state predicted from the T-matrix solution and the result of our selfconsistent procedure exists. This shift stems from the fact that the T-matrix solution only applies exactly for a purely $d_{z^2}$ impurity $V_0^{\mu} = \delta_{\mu, z^2}V_0$ and neglects the effect of selfconsistent density modulations, while the CBdG result includes a multiorbital impurity and selfconsistently converged the mean fields. Repeating the CBdG procedure without selfconsistency and assuming a pure $d_{z^2}$ impurity exactly reproduces the expected resonant state positions as seen in \ref{fig:BS_variation} \fl{d}. 
The close correspondence between the T-matrix predictions of resonant states and the obtained phase diagrams indicate that these regions of local order can be efficiently obtained by first considering the homogeneous Greens function. The search of pockets of local magnetic order in ($U,V_0$) space is thus made much simpler as approximate phase diagrams can be obtained from a single calculation in the clean system.

\begin{figure}[t!]
\includegraphics[width = \linewidth]{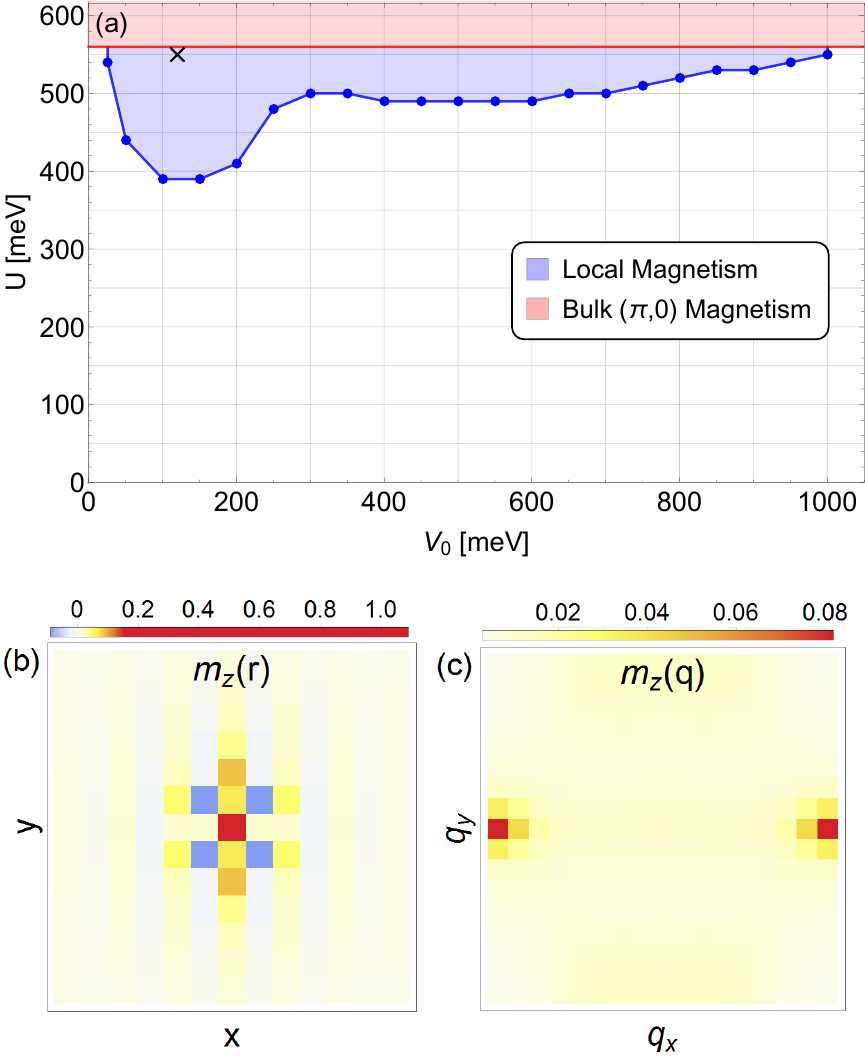}
	\caption{
	\fl{a} Phase diagram of the orbital-selective model showing the transition to a bulk $(\pi, 0)$ phase above $U_c = 560$ meV (red), and the region of local magnetic order (blue) as a function of the impurity potential strength $V_0$ and interaction $U$. Apart from shifting the phase boundaries, the orbitally selective interaction parameters also fundamentally alter the bulk and local magnetic orderings.   
	\fl{b} Zoom of the local magnetic structure nucleated around the impurity site for potentials close to the bulk transition, $U = 550$ meV and $V = 120$meV (black cross in the phase diagram). \fl{c} Fourier transform of the local magnetic order  revealing the highly anisotropic local $(\pi,0)$ structure.
	 }
	 \label{fig:PD_OS}
\end{figure}

\section{Effects of orbital selectivity}
The physics of orbital selectivity has been studied quite extensively in correlated multi-orbital models relevant for FeSCs\cite{Biermann_review,Guterding2017,deMedici_review}. In particular several groups have applied DMFT\cite{Haule2009,Yin2011} and slave-spin methods\cite{deMedici2014,deMedici2011} to investigate self-energy effects on e.g. the band-structure. Such studies have found strong orbital dependent mass renormalizations and quasi-particle weights $Z_\mu$. Motivated by the recent experimental evidence for orbital selectivity in FeSe, we construct also a ``dressed'' version of the above mean-field model. This can be done most simply by the prescription
\begin{align}
c_{i\mu}^\dagger &\to \sqrt{Z_\mu} c_{i\mu}^\dagger,  \label{eq:ansatz}
\end{align}
where $Z_{\mu}$ denotes the quasi-particle weight for the orbital $\mu$. The orbitally selective ansatz in Eq. (\ref{eq:ansatz}) leads to a modified mean-field theory where all effects of orbital selectivity are contained in dressed interaction parameters 
\begin{align}
U_{\mu}   &\to Z_{\mu}^2 U_{\mu}, \\
U'_{\mu \nu}  &\to Z_{\mu} Z_{\nu} U'_{\mu \nu}, 
\end{align}
with similar expressions for $J, J'$. Based on earlier studies of FeSe, in the following we fix the values of the quasi-particle weights
 $\lbrace \sqrt{Z_{\mu}} \rbrace = \lbrace 0.2715, 0.9717, 0.4048, 0.9236, 0.5916 \rbrace $ for the five Fe 3d orbitals $\lbrace d_{xy}, d_{x^2-y^2}, d_{xz}, d_{yz}, d_{z^2} \rbrace$. We note that these values are within the confidence interval of the experimentally extracted values of $Z_{\mu}$.\cite{Kreisel2017,Kreisel_2018arXiv180709482K}

Making this orbitally selective ansatz and including the quasi-particle weights $Z_{\mu}$ defined above changes the magnetic phase diagram as shown in Fig. \ref{fig:PD_OS} \fl{a}. In Ref. \onlinecite{Kreisel2017} the splitting of $Z_{yz}, Z_{xz}$ quasiparticle weights was shown to result in a leading $(\pi,0)$ stripe order instability, in agreement with the $(\pi,0)$ ordered bulk phase of our selfconsistent calculations. Close to the phase transition we again find local magnetic order as displayed in Fig. \ref{fig:PD_OS} \fl{b}, which inherits the bulk $(\pi,0)$ structure as seen in the Fourier transform of the magnetization in \fl{c} that exhibits only a peak at $(\pi,0)$ and is strongly $C_2$ symmetric.  We find the ordering structure of this local magnetism to vary with the Hubbard $U$, starting out strongly $C_2$-symmetric just below the bulk phase transition, and then transitioning to a nearly $C_4$-symmetric $(\pi,0) + (0,\pi)$ structure when approaching the lower local order boundary line. This result is in sharp contrast to the omnipresent nearly $C_4$-symmetric magnetization exhibited by the bare model. Similar to the results from the bare model, we find that this region of local magnetic order can be understood from the emergence of resonant states just below the transition.

\section{Detection and differentiation of local magnetism}

\begin{figure}[tb]
\includegraphics[width = \linewidth]{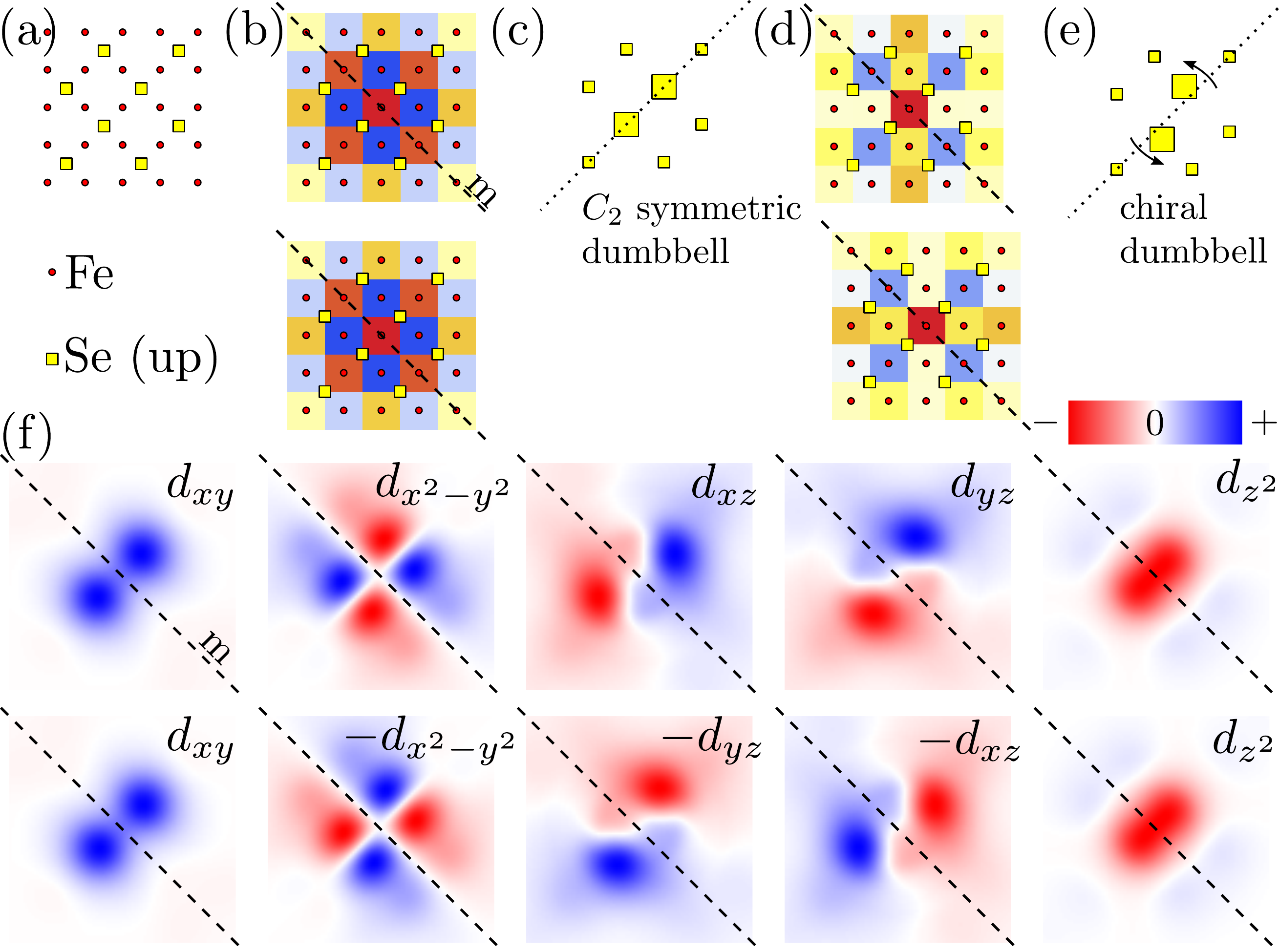}
\caption{Symmetries of the order parameter and the Wannier functions on the surface of FeSe.
\fl a Positions of the Fe atoms (red circles) and the Se atoms (yellow squares) at the surface of FeSe.
\fl b Magnetic order parameter around an impurity for the case without orbital selectivity which exhibits an (approximate) symmetry for a mirror plane $m$ along the diagonals (dashed lines).
\fl c Expected pattern of the local density of states at the STM tip position for case (b).
\fl d Magnetic order parameter in the orbital selective case which does not exhibit the mirror symmetry such that the expected pattern in an STM experiment shows deviations of the maxima from the symmetry axis (dotted line) \fl e.
\fl f Cuts through the five Wannier functions for FeSe\cite{Choubey2014,Kreisel_16} (red/blue: negative/positive) centered at one Fe atom which have definite symmetry properties with respect to mirror plane $m$ (lower row: mirror operation applied to function).}
     \label{fig:Symmetry_Wannier}
\end{figure}

Specifically for the material FeSe, there exists evidence for local magnetism from bulk experimental probes as discussed in the introduction.\cite{He2018,Grinenko2018,Khasanov2008} However, given the high quality of the available crystals\cite{Boehmer2013}, local probes like STM may be more suitable for direct investigation of the nature of the electronic state in the vicinity of impurities. For the appearance of static local magnetic order, an obvious experimental technique would be spin-polarized STM measurements. While such approaches have been carried out and are developed recently, it is worth pointing out that also a non-spin polarized experiment can be used to discriminate between the two scenarios of local magnetization which we have investigated. At this point, we do not intend to perform a quantitative simulation of topographies and conductance maps,\cite{Choubey2014,Kreisel_16} but instead utilize simple symmetry based arguments that hold true also in the case of a correlated electron system. In order to calculate the tunneling current as measured in STM experiments, one needs to consider the LDOS at the position of the STM tip.\cite{Tersoff85}
If the underlying model Hamiltonian is constructed on a lattice, as in the present case, see Eq.~(\ref{eq:H_0}), the connection to the relevant quantity above the surface of the material, i.e. in the vacuum, can be made by a basis transformation where Wannier functions of the electronic states enter as matrix elements.\cite{Choubey2014,Kreisel_16} For single impurities, it has been shown that the properties of the elementary cell have imprints on the observed shapes in topographies and conductance maps\cite{Choubey2014,Kreisel_16, Kreisel_Wannier_15, Chi_16, Huang2017}. In Fig. \ref{fig:Symmetry_Wannier} (a), we show the positions of the atoms at a cleaved surface of FeSe, where the Fe atoms form the lattice as used in our Hamiltonian, while the Se atoms above the Fe plane are arranged in a rotated lattice with larger lattice constant which is also observed in STM experiments\cite{Sprau2017,Kostin2018,Li2017}. For the case of FeSe, a Fe centered impurity, leads to the observation of a dumbbell originating from the tails of the Wannier functions that have weights close the positions of the Se atoms at the surface of the material.\cite{Choubey2014}

Now, let us turn to the symmetries of the magnetic order parameter as presented in Figs. \ref{fig:PD_0} and \ref{fig:PD_OS} by considering a mirror plane $m$ along the diagonals of the Fe lattice, as shown by the dashed line in Fig. \ref{fig:Symmetry_Wannier}. The Wannier functions of the five relevant states at one Fe atom, of course have definite symmetry properties under this operation which is shown explicitly in Fig. \ref{fig:Symmetry_Wannier} (f) where in the top row, maps of Wannier functions in FeSe above the surface, i.e. at the STM tip are presented and in the lower row, a mirror operation has been applied. First, we note that the shape of the wavefunctions have lower symmetry than the corresponding atomic wavefunctions and second, the $d_{xz}$ and $d_{yz}$ Wannier functions exhibit a chiral structure of opposite direction. Turning now to the two patterns of local magnetic order, we see that for the order of $(\pi,\pi)$ type, Fig. \ref{fig:Symmetry_Wannier} (b), the mirror plane is a symmetry and therefore the expected pattern in an STM experiment will be symmetric with respect to the mirror plane $m$ as well, such that one should expect that enhancements at Se positions should be along the diagonal (dotted line) as presented schematically in Fig. \ref{fig:Symmetry_Wannier} (c)\footnote{Strictly speaking, the magnetization profile in Fig. 4(c) breaks $C_4$ symmetry since it is generated from the nematic FeSe bandstructure. This, however, is a tiny effect not important for the general arguments for differentiation between different magnetization patterns by STM.}. As for the chiral components of the $d_{xz}$ and $d_{yz}$ Wannier functions, these will enter with equal weights in the calculation of the LDOS such that the final pattern does not exhibit any chiral character. This situation is different for the local magnetic order parameter of $(\pi,0)$ type which obviously does not have a definite symmetry under the mirror operation $m$, Fig. \ref{fig:Symmetry_Wannier} (d) and therefore should produce a pattern in the LDOS where the maxima are away from the dotted line in Fig. \ref{fig:Symmetry_Wannier} (e). Similar features have been found in bulk FeSe\cite{Kostin2018} and have been recently analyzed quantitatively in experiments reporting local impurity-induced magnetization in thin films of FeSe\cite{Li2017}. The direction of the deviation depends on the details of the orbital structure of the local order parameter, and could also switch as a function of bias voltage.

 \section{Conclusions}

In summary, we have explored theoretically the induction of local static magnetic order by nonmagnetic impurity potentials in a model relevant for FeSe. We have mapped out the regions of the phase diagram where such order is present, and investigated the role of orbital-selectivity. The latter may strongly alter the magnetic structure of both long-range and short-range magnetism. Finally we discussed the detection of local magnetic order by non-spin-polarized STM measurements and provided simple symmetry-based arguments to illustrate how this technique may be used to differentiate between distinct forms of induced local magnetic order. 

\section{Acknowledgements}

We acknowledge useful discussions with M. N. Gastiasoro, P. J. Hirschfeld, and Daniel D. Scherer. The Center for Nanostructured Graphene is supported by the Danish National Research Foundation, Project DNRF103. 


%

\end{document}